\documentclass{iau} 
\usepackage{graphicx}
\usepackage{amsmath}	
\usepackage{amssymb}	
\usepackage[version=3]{mhchem}
\usepackage{subfigure}

\title[On the origin of volatile species in comets] 
{On the origin of O$_2$ and other volatile species in comets}

\author[Vianney Taquet, Kenji Furuya, Catherine Walsh \& Ewine
F. van Dishoeck]   
{Vianney Taquet$^{1,2}$,
Kenji Furuya$^{1,3}$,
Catherine Walsh$^{1,4}$, \and
Ewine F. van Dishoeck$^{1,5}$}

\affiliation{$^1$Leiden Observatory, Leiden University, PO Box 9513, 2300 RA, Leiden, The Netherlands \\ 
$^2$INAF-Osservatorio Astrofisico di Arcetri, Largo E. Fermi 5, I-50125 Firenze, Italy \\
$^3$Center for Computational Sciences, University of Tsukuba, 1-1-1 Tennoudai, 305-8577, Tsukuba, Japan \\
$^4$School of Physics and Astronomy, University of Leeds, Leeds LS2 9JT, UK \\
$^5$Max-Planck-Institut f{u}r Extraterrestrische Physik, Giessenbachstrasse, 85741, Garching, Germany
}

\pubyear{2017}
\volume{332}  
\jname{Astrochemistry VII -€" Through the Cosmos from Galaxies to Planets.}
\editors{M. Cunningham, T. Millar \& Y. Aikawa, eds.}
\begin{document}

\maketitle

\begin{abstract}
Molecular oxygen, O$_2$, was recently detected in comet 67P by the ROSINA
instrument on board the Rosetta spacecraft with a surprisingly high
abundance of 4\% relative to H$_2$O, making O$_2$ the fourth most
abundant in comet 67P. Other volatile species with similar volatility,
such as molecular nitrogen N$_2$, were also detected by Rosetta, but
with much lower abundances and much weaker correlations with water. 
Here, we investigate the chemical and physical origin of O$_2$ and
other volatile species using the new constraints  
provided by Rosetta. We follow the chemical evolution during star
formation with state-of-the-art astrochemical models applied to
dynamical physical models by considering three origins: i) in dark
clouds, ii) during forming protostellar disks, and iii) during
luminosity outbursts in disks. The models presented here favour a dark
cloud (or ``primordial'') grain surface chemistry origin for  volatile species in comets,
albeit for dark clouds which are slightly warmer and denser than those usually
considered as solar system progenitors.  

\keywords{astrochemistry, comets: individual: 67P/C-G, ISM: abundances, ISM:
molecules, protoplanetary discs , stars: formation}
\end{abstract}

\firstsection 
\section{Introduction}

The {\it Rosetta} spacecraft analysed
the Jupiter-family comet 67P/Churyumov-Gerasimenko (hereinafter comet
67P/C-G) in 2014 and 2015. The ROSINA instrument on board the
{\it Rosetta} orbiter (Rosetta Orbiter Spectrometer for Ion and
Neutral Analysis (\cite[Balsiger et al. 2007]{balsiger2007}) detected
a ``zoo'' of molecules in the coma of 67P/C-G from simple mono-atomic
or di-atomic species to complex pre-biotic molecules, such as glycine,
the simplest amino-acid \cite[(Le Roy et al. 2015, Altwegg et
al. 2016)]{leroy2015, altwegg2016}.  

One of the most surprising results provided by the ROSINA instrument is
the {\it in-situ} detection of molecular oxygen, \ce{O2}, in the coma of
comet 67P/C-G, resulting in the first detection of \ce{O2} in a comet
(\cite[Bieler et al. 2015]{bieler2015}). 
\ce{O2} is strongly correlated with \ce{H2O} and is present at an
average level of $3.80 \pm 0.85$\% relative to \ce{H2O}, making it the
fourth most abundant molecule in the comet,  following \ce{H2O},
\ce{CO2}, and \ce{CO}.   
The authors argue that \ce{O2} does not originate from gas-phase
chemistry in the coma but from direct sublimation from or within the
comet surface.   
Moreover, the strong correlation with \ce{H2O} suggests that the \ce{O2} 
is trapped within the bulk \ce{H2O} ice matrix of the comet and
therefore that \ce{O2} was present within the ice mantle on dust grains
in the presolar nebula prior to comet formation since the surface of
comet 67P/C-G revealed today is likely pristine.  
%
A reanalysis of data from the Neutral Mass Spectrometer on board 
the {\em Giotto} probe which did a fly-by of comet 1P/Halley in 1986, 
confirmed the presence of \ce{O2} at a level similar to that seen in
67P/C-G (\cite[Rubin et al. 2015b]{rubin2015b}).   
This suggests that \ce{O2} is not only an abundant molecule in comets,
but is also common to both Jupiter-family comets,  such as 67P/C-G,
and Oort Cloud comets, such as 1P/Halley, which have different
dynamical behaviours and histories.     
These results raise the question whether \ce{O2} was abundant 
in icy dust mantles entering the protoplanetary disk of the young Sun, 
or whether the conditions in the comet-forming zone of the early solar system
were favourable for \ce{O2} formation and survival.  

\ce{O2} is a diatomic homonuclear molecule; hence it does not possess
electric dipole-allowed rotational transitions. Therefore,
gas-phase \ce{O2} has been particularly elusive in interstellar
clouds. 
Recent high sensitivity observations with the {\em Herschel Space Observatory} 
allowed a deep search for \ce{O2} towards sources considered as true
solar system progenitors: low-mass protostars.     
A deep upper limit was determined towards the well-studied 
protostar, NGC~1333-IRAS~4A, (\ce{O2}/\ce{H2} $\leq
6\times10^{-9}$ or \ce{O2}/\ce{H2O}  $\leq 0.012$\% with a
\ce{H2O} abundance of $\sim 5 \times 10^{-5}$, \cite[Yildiz et
al. 2013]{yildiz2013}).    
This picture is consistent with laboratory experiments that 
have shown that \ce{O2} ice is efficiently hydrogenated at 
low temperatures and converted into \ce{H2O} and \ce{H2O2} ices
(\cite[Ioppolo et al. 2008, Miyauchi et
al. 2008]{ioppolo2008,miyauchi2008}).   
This makes the close association of \ce{O2} with \ce{H2O} in 67P/C-G 
an even stronger enigma.  

Despite \ce{O2} being a particularly elusive molecule in interstellar 
and circumstellar environments, there apparently do exist conditions which 
are favourable for the formation of \ce{O2} and related species 
at abundance ratios similar to that observed in ices in comet 67P/C-G.  
Indeed, {\em Herschel} did reveal the presence of gas-phase \ce{O2}
in two sources:  
an active star forming region in Orion (\cite[Goldsmith et al. 2011]{goldsmith2011})
and in the dense core $\rho$~Oph~A located in the more quiescent 
$\rho$~Oph molecular cloud complex, which stands out  
from other low-mass star-forming regions by exhibiting emission from relatively 
warm molecular gas (\cite[Larsson et al. 2007, Liseau et
al. 2012]{larsson2007, liseau2012}).     
Subsequent observations of $\rho$~Oph~A have also determined the presence 
of related gas-phase species, \ce{HO2} and \ce{H2O2}, at abundance
levels in reasonable agreement with those seen in 67P/C-G with ROSINA
($\sim 2\times10^{-3}$ that of \ce{O2}, see \cite[Bergman et al. 2011b,
Parise et al. 2012]{bergman2011b,parise2012}).   
The chemically related species, \ce{O3} (ozone), was not detected in the comet coma 
with a very low upper limit, $< 2.5\times 10^{-5}$ with respect to
\ce{O2}. 

Other key di-atomic molecules of similar volatility, CO and \ce{N2}, have also been
detected in 67P/C-G by {\it Rosetta} but with a much weaker
correlation with H$_2$O (\cite[Rubin et al. 2015a, Bieler et
al. 2015]{rubin2015a, bieler2015}). Although CO shows a high abundance
of 10-30\% relative to \ce{H2O}, in good agreement with
previous observations towards other comets, \ce{N2} shows a much lower
abundance of $0.57 \pm 0.07$ \% relative to CO. The different
correlations and abundances w.r.t. the \ce{H2O} clearly suggest
a different chemical history for \ce{O2}, CO, and \ce{N2}.

Here we explore and discuss several different origins to explain the
strong constraints provided by {\it Rosetta} on \ce{O2} and other
volatile species in comet 67P/C-G:
i) in dark clouds (``primordial'' origin), ii) during the journey from
the protostellar envelope into the disk, iii) during luminosity
outbursts within the protoplanetary disk.

\section{Astrochemical models}


The gas-grain astrochemical models by \cite{taquet2014} and
\cite{furuya2015} have been used 
in this work to study the formation and survival of \ce{O2} and other
volatile species from dark clouds to the Solar System.  
These models couple the gas phase and ice chemistries with the approach developed by
\cite{hasegawa1993} to follow the multi-layer formation of
interstellar ices and to determine the gas-ice balance.  
Several sets of differential equations, one for gas-phase species, one
for surface ice-mantle species, and one (or several) for bulk
ice-mantle species, are considered to follow the time-evolution of
abundances.
Following \cite{vasyunin2013}, the chemically-active surface is 
limited to the top four monolayers.  
The original three-phase model considered in the Taquet model assumes
that the inert bulk ice mantle  has a uniform molecular composition.
In order to accurately follow the ice evolution in warm conditions,
the Furuya model considers a depth-dependent molecular 
composition, through the division of the 
inert bulk ice mantle into five distinct phases (for details,
see \cite[Furuya et al. 2016]{furuya2016} and references therein). 

The gas-phase chemical network used by the Taquet model is based on
the 2013 version of the KIDA chemical database 
\cite[(Wakelam et al. 2012)]{wakelam2012}. It has been further updated
to include warm gas-phase chemistry involving water and and
ion-neutral reactions involving ozone.  
The network also includes the surface chemistry of all 
dominant ice components,  as well as those important for water (e.g.,
\ce{O2}, \ce{O3}, and \ce{H2O2}).   
Several new surface reactions were added involving \ce{O3} and reactive species 
such as N, O, OH, \ce{NH2}, and \ce{CH3}, following the NIST gas-phase
chemical database.
The gas-ice chemical network of \cite{garrod2006}, based on the OSU
2006 network, is used in the Furuya model.
 The gas phase and surface networks in the Furuya model are more suited to the high
 density and warm temperatures conditions found in 
protostellar envelopes. It has therefore been supplemented with
high-temperature gas-phase reactions from \cite{harada2010} and includes the
formation of many complex organic molecules. It is consequently more
expansive than the network used in the Taquet model. 

Elemental abundances of species used in the two models correspond to
the set EA1 from \cite{wakelam2008}.  
Standard input parameters assumed for the two astrochemical models
are: a cosmic ray ionisation rate $\zeta$ of $1 \times 10^{-17}$ s$^{-1}$, a
flux of secondary UV photons of $10^4$ phot. cm$^{-2}$ s$^{-1}$, a
dust-to-gas mass ratio of 1\%, a grain diameter of 0.2 $\mu$m, a
volumic mass of grains of 3 g cm$^{-3}$, a grain surface density of 10$^{15}$ cm$^{-2}$, a
diffusion-to-binding energy ratio of 0.5, four chemically active
monolayers, and a  sticking coefficient of species heavier than H and
\ce{H2} of 1.

\section{Interstellar chemistry of molecular oxygen}

Two main processes have been invoked for the formation of
molecular oxygen in the interstellar medium: i) gas-phase formation 
via neutral-neutral chemistry, and
ii) formation via association reactions on/within icy mantles of dust
grains. 
Gaseous \ce{O2} is thought to form primarily via the barrierless
neutral-neutral reaction between O and OH in cold and warm gas.
Due to its importance, this reaction has been well studied 
both experimentally and theoretically.  
The formation of \ce{O2} in cold dark clouds is initiated by the high
initial abundance assumed for atomic oxygen, 
inducing an efficient ion-neutral chemistry that also forms OH. 
In warm environments ($T\gtrsim100$~K), e.g., the inner regions of 
protostellar envelopes or the inner, warm layers of protoplanetary disks, 
OH and O are mostly produced through warm neutral-neutral chemistry driven 
by the photodissociation of water sublimated from interstellar ices. 
Solid \ce{O2} in dark clouds is involved in the surface
chemistry reaction network leading to the formation of water ice
\cite[(Tielens \& Hagen 1982, Miyauchi et al. 2008, Ioppolo et
al. 2008)]{tielens1982,miyauchi2008, ioppolo2008}.   
\ce{O2} is formed through atomic O recombination on ices and 
efficiently reacts with either atomic O or atomic H to form \ce{O3} or
\ce{HO2}, respectively, eventually leading to the formation of
water. 
The hydrogenation of \ce{O3} also leads to the formation of \ce{O2},  
in addition to dominating the destruction of \ce{O3}.

Radiolysis, i.e. the bombardment of (ionising) energetic particles depositing energy
into the ice, and/or photolysis, i.e. the irradiation of ultraviolet
photons breaking bonds, can trigger chemistry within
the mantle of cold interstellar ices. 
We have investigated the impact of the UV photolysis induced by secondary
UV-photons on the bulk ice chemistry and the formation and survival of
\ce{O2}. We find that \ce{O2} cannot be efficiently produced in the bulk
through ice photolysis as the photodissocation of the main ice
components not only produces O atoms, that recombine together to
form \ce{O2}, but also H atoms that react with \ce{O2} to reform
water.
Laboratory experiments show that \ce{O2} can be efficiently formed through
radiolysis of ices without overproducing \ce{H2O2} only if the radiolysis
occurs as water is condensing onto a surface (see \cite[Teolis et al. 2006]{teolis2006}). 
However, in dark clouds  water ice is mostly formed {\it in-situ} at the
surface of interstellar grains through surface reactions involving
hydrogen and oxygen atoms.  This happens prior to the formation of the
presolar nebula, i.e. the cloud out of which our solar system was
formed, and it is possible that the comet-forming zone of the 
Sun's protoplanetary disk inherited much of its water ice from the
interstellar phase \cite[(Visser et al. 2009, Cleeves et
al. 2014)]{visser2009, cleeves2014}.

\section{Origin of cometary \ce{O2}}

\subsection{Dark cloud origin?}

{\underline{\it Impact of physical and chemical parameters}}. 
We first investigated whether the \ce{O2} observed in 67P/C-G
has a dark cloud origin. For this purpose, we used the Taquet
astrochemical model presented in section 2.
We carried out a first parameter study, in which 
several surface and chemical parameters are varied, in order to 
reproduce the low abundances of the chemically related species
\ce{O3}, \ce{HO2}, and \ce{H2O2} with respect to \ce{O2} seen in comet
67P/C-G. The low abundance of \ce{O3} and \ce{HO2} relative to
\ce{O2} ($\leq 2 \times 10^{-3}$) can be explained when a small activation barrier of $\sim$ 300
K is introduced for the reactions O~+~\ce{O2} and H~+~\ce{O2}, in
agreement with the Monte-Carlo modelling of laboratory experiments by
\cite{lamberts2013}. However, the abundance of \ce{H2O2} is still
overproduced by one order of magnitude, suggesting that other chemical
processes might be at work.  

A second parameter-space study was then conducted to determine the
range of physical conditions over which \ce{O2} ice and gas  (and
those for chemically-related species, \ce{O3}, \ce{HO2}, and
\ce{H2O2}) reach abundances (relative to water ice) similar to that seen in
67P/C-G.  
We ran a model grid in which four or five values for the total density of H
nuclei, $n_{\textrm{H}}$, the gas and dust temperature, $T$ (assumed
to be equal), the cosmic ray ionisation rate, $\zeta$, and the visual
extinction, $A_{\textrm{V}}$ are considered, following the methodology 
described in \cite{taquet2012}, resulting in 500 models in total.  
We explored the distribution of abundances of solid
\ce{O2}, and the chemically related species, \ce{O3}, \ce{HO2}, and \ce{H2O2}, 
relative to water ice, when the time reaches the free-fall time, $t_\mathrm{FF}$. 
The results show that the formation and survival of solid \ce{O2}, and
other reactive species, in interstellar ices, is  
strongly dependent upon the assumed physical conditions with
abundance distributions ranging over several orders of magnitude.  
High \ce{O2} abundances ($\geq 4$\% relative to water ice) are obtained only for the 
models with high densities ($n_{\textrm{H}} \gtrsim 10^5$ cm$^{-3}$).  
Higher gas densities result in a lower gas-phase H/O ratio, thereby
increasing the rate of the association reaction between  O atoms to
form \ce{O2} ice, and correspondingly decreasing the rate of the competing
hydrogenation reactions, O~+~H and \ce{O2}~+~H,  which destroy \ce{O2}
ice once formed.     
An intermediate temperature of 20~K is also favoured because it
enhances the mobility of oxygen atoms on the grain surfaces whilst at the same time 
allowing efficient sublimation of atomic H.   
This additionally enhances the rate of oxygen recombination 
forming \ce{O2}, with respect to the competing hydrogenation
reactions.
Moreover, because the density of gas-phase H atoms increases linearly with the
cosmic-ray ionisation rate, $\zeta$, a low value of $\zeta$ also tends to
favour the survival of \ce{O2} ice. 
On the other hand, the visual extinction does not have a
strong impact on the abundance of solid \ce{O2}.

\begin{figure}
\centering 
\includegraphics[width=0.25\columnwidth]{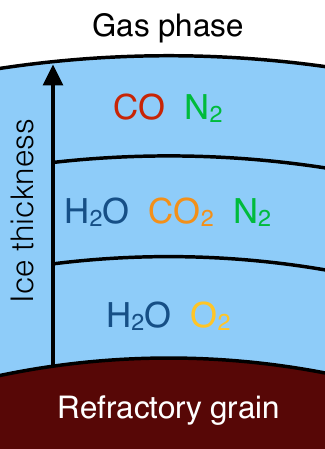}
\includegraphics[width=0.35\columnwidth]{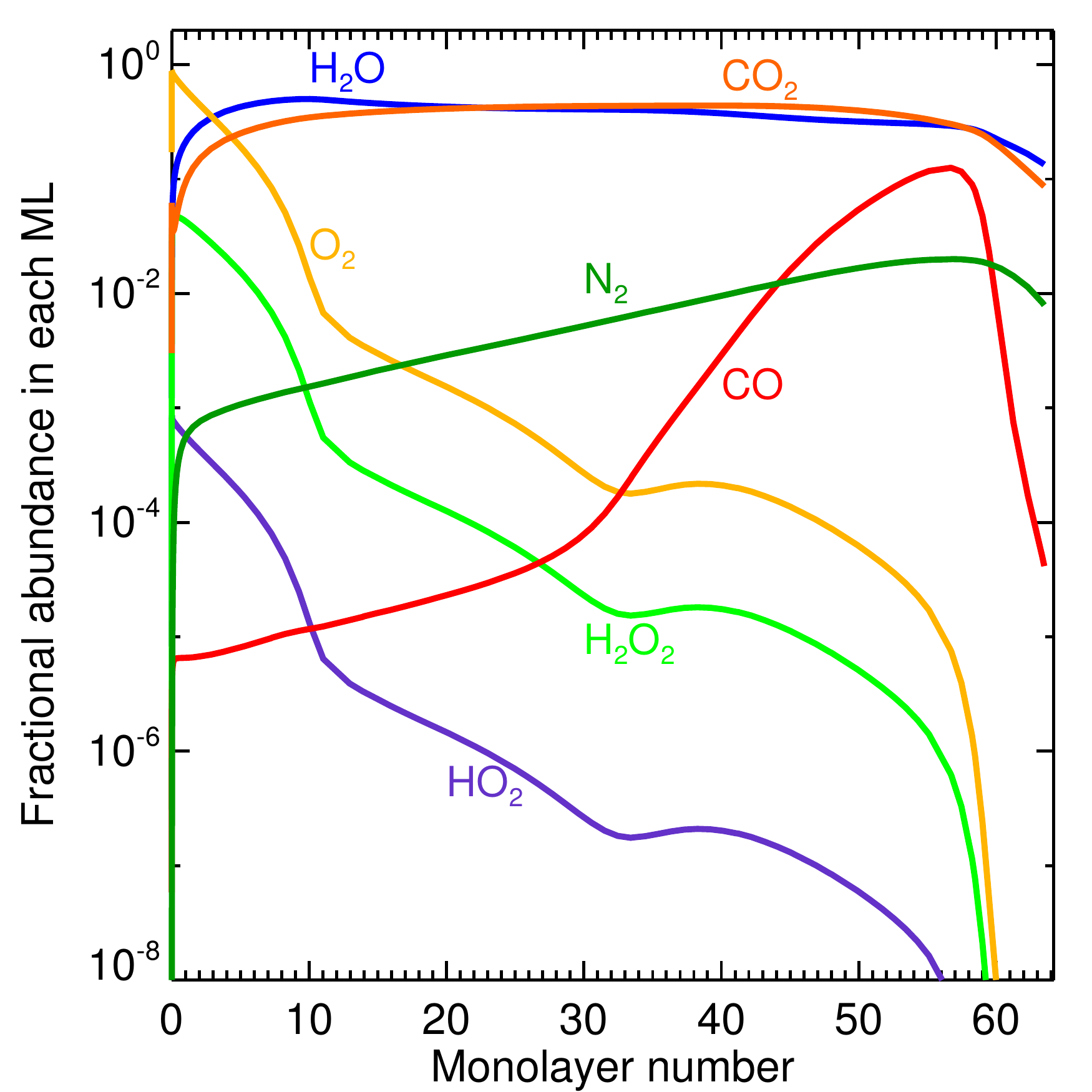}
\includegraphics[width=0.35\columnwidth]{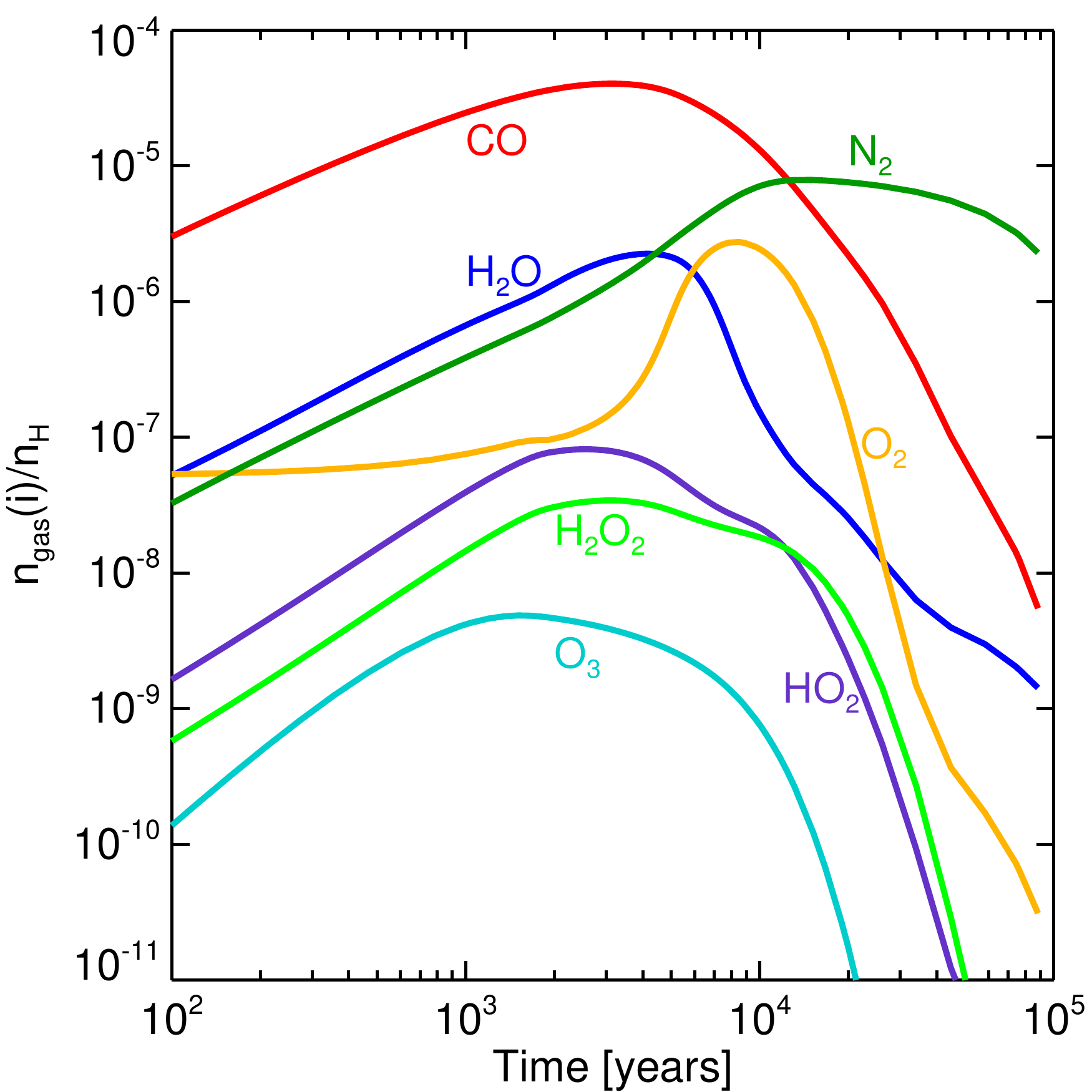} 
\caption{
Left panel: Cartoon representation of the ice structure predicted in
this work. 
Middle panel: Fractional composition of each ice monolayer as function of
  the monolayer number or ice thickness.
Right panel: gas phase
  abundances as a function of time of \ce{O2} and its chemically related
  species predicted by the model using the   $\rho$ Oph A physical conditions.
}
\label{rhoopha}
\end{figure}

{\underline{\it The $\rho$~Oph~A case}}. 
The parameter study presented above therefore suggests that the physical conditions of
$\rho$~Oph~A, presenting a high density ($n_{\textrm{H}} \sim 10^6$
cm$^{-3}$), and  a relatively warm temperature for a starless core
($T_{\textrm{kin}} = 24 - 30$ K and $T_{\textrm{dust}} \sim 20$ K; \cite[Bergman et
al. 2011a]{bergman2011a}) are consistent with those which facilitate
the formation and survival of \ce{O2} ice. 
This confirms that these properties offer optimal conditions for an efficient
production of solid \ce{O2} since $\rho$~Oph~A is the only
interstellar source so far where gas-phase \ce{O2}, \ce{HO2}, and
\ce{H2O2} have been detected.  
Figure \ref{rhoopha} shows the chemical composition of the ice and gas
obtained for the model using the physical conditions of  $\rho$~Oph~A
and that best reproduce the observations in comet 67P/C-G.
The fractional composition in each ice monolayer is plotted as
function of monolayer number,  i.e. the ice thickness that grows with time. 
\ce{O2} ice is mostly present in the innermost layers of the ice mantle
and decreases in relative abundance towards the ice surface,
reflecting the initial low ratio of H/O in the gas phase obtained at
high densities, but tends to be well mixed with \ce{H2O} ice.  
\ce{CO2} is also highly abundant because the higher temperature (21~K)
enhances the mobility of heavier species, such as \ce{O} or CO.

\ce{O2}, \ce{O3}, \ce{HO2}, and \ce{H2O2} are mostly, and potentially
only, produced via surface chemistry; hence their gas-phase abundances depend on their
formation efficiency  in interstellar ices and on the probability of
desorption upon formation through chemical desorption (thought to be
the dominant desorption mechanism for these species in dark cloud conditions).   
The chemical desorption probabilities are highly uncertain and mainly
depend on the ice substrate and the considered reaction. 
Figure \ref{rhoopha} shows the temporal evolution of the gas
phase abundances of \ce{O2}, \ce{O3}, \ce{HO2}, and \ce{H2O2} when the
theoretical values by \cite{minissale2016} relative to a bare grain
substrate, and varying between 0 and 70\%, are used.
This model is almost able to simultaneously reproduce the
gaseous abundances of \ce{O2}, \ce{HO2}, and \ce{H2O2} derived in $\rho$ Oph A
since the predicted \ce{O2} abundance and the \ce{HO2} and \ce{H2O2}
abundances reach the observations at similar timescales ($1.5 \times
10^4$ vs $2.2 \times 10^4$ yr). Using lower chemical desorption
probabilities relevant to water ice substrates could improve the
comparison with the observations.

{\underline{\it Molecular nitrogen vs carbon monoxide and molecular oxygen}}.
In contrast to \ce{O2}, it is seen that \ce{CO} and \ce{N2} are mostly formed in
the outer part of the ices and would, therefore, undergo a more efficient
sublimation, either thermally or through photo-evaporation, during
their transport from dark clouds to forming disks in the subsequent
protostellar collapse phase. 
The chemical heterogeneity predicted in ices can therefore naturally
explain the high correlation between \ce{O2} and \ce{H2O} signals together
with the weak correlation between CO, \ce{N2}, and \ce{H2O} signals
measured in comet 67P/C-G. However, it cannot explain the low
\ce{N2}/CO abundance ratio of $\sim 0.6$ \% observed in comet
67P/C-G since our dark cloud model predicts a \ce{N2}/CO of 50 \%. 

As shown by dynamical models of protoplanetary disk formation,
volatile species that evaporated during their journey from dark
clouds to upper disk layers can subsequently freeze-out onto ices again
once they reach the colder disk midplane (\cite[see Drozdovskaya et
al. 2014]{drozdovskaya2014}). 
\ce{N2} is known to be slightly more volatile than CO with a
binding energy lower by $\sim 150$ K for H$_2$O ice and by $\sim 60$ K
for pure ices (\cite[Bisschop et al. 2006, Fayolle et
al. 2016]{bisschop2006, fayolle2016}). 
We investigated the impact of these slightly different binding
energies with a toy model on the recondensation of CO and \ce{N2}
during a cooling from 50 to 20 K that could occur during the
transport of material from the upper disk layers to the disk
midplane, assuming a constant density typical of a disk ($n_{\rm H} =
10^8$ cm$^{-3}$) and that all CO and \ce{N2} are initially in the gas
phase.
It is found that CO does freeze out more efficiently than \ce{N2},
inducing a low \ce{N2}/CO abundance ratio in ices down to 0.2\% at
28 K before a re-increase to the initial abundance ratio at lower
temperatures.  
A cooling of ices down to 26 - 28 K near the cometary zone therefore induces a
CO/\ce{H2O} abundance higher than 10\% and a \ce{N2}/CO abundance
lower than 0.6\%, even if a high initial \ce{N2}/\ce{CO} abundance is
produced in the dark cloud phase.
These numbers are close to those observed for 67P/C-G.

\subsection{Disk formation origin?}

Here, we discuss the role of chemistry during protostellar collapse and 
protoplanetary disk formation on the observed abundance of \ce{O2} in
67P/C-G.   
To follow the chemical evolution from prestellar cores to forming
disks, fluid parcels from the envelope to the disk are traced with the
physical model initially developed by \cite{visser2009}.
The Furuya astrochemical model is used to follow the gas-ice
chemical evolution calculated along each individual trajectory. 
The physical model used here is an axisymmetric semi-analytical
two-dimensional model that describes the temporal evolution of the
density and  velocity fields following inside-out collapse and the
formation of an accretion disk described by the $\alpha$-viscosity
prescription (for details, see \cite[Harsono et al. 2013]{harsono2013} and references therein).
The model follows the physical evolution until the end of the main 
accretion phase when the gas accretion from the envelope 
onto the star-disk system is almost complete.
A molecular cloud formation model is run to determine the 
composition of the gas and ice in the parent molecular cloud 
(\cite[Furuya et al. 2015]{furuya2015}).  
The chemistry is then evolved for an additional $3 \times 10^5$ yr 
under prestellar core conditions to compute the abundances at the onset of 
collapse. 
At the onset of collapse, models have a negligible \ce{O2} ice abundance.  

\begin{figure*}
\centering
\includegraphics[width=\textwidth]{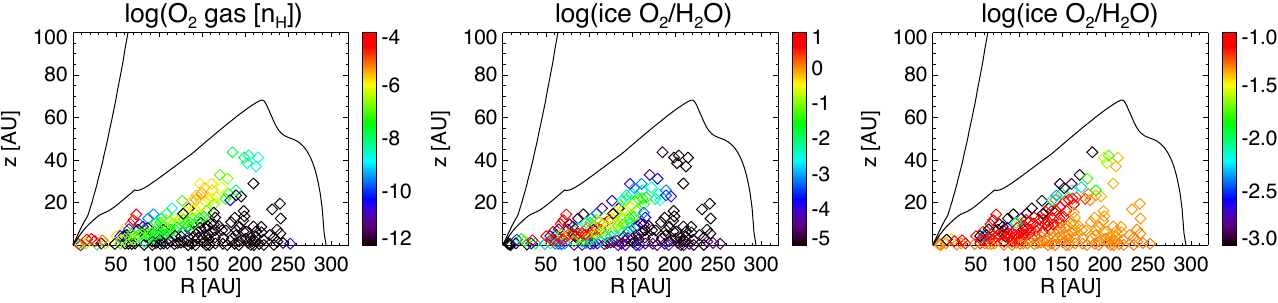}
\caption{Spatial distributions of fluid parcels at the final time of
  the spreak-dominated disk formation simulation ($\Omega=10^{-13}$ s$^{-1}$).
The left panel shows the gaseous \ce{O2} abundance with respect to hydrogen nuclei,
the middle panel shows the abundance ratio between \ce{O2} ice and \ce{H2O} ice.
the right panel also shows the abundance ratio between \ce{O2} ice and \ce{H2O} ice,
but for the model where the initial ratio is artificially set to 5\%.
The solid lines represent the outflow cavity wall and the disk surface.}
\label{fig:disk_formation1}
\end{figure*}

Figure \ref{fig:disk_formation1} shows the spatial distributions of 
fluid parcels at the final time of the simulation 
in models in the so-called spread-dominated case (initial core
rotation rate $\Omega = 10^{-13}$ s$^{-1}$). 
It is found that
(i) some gaseous \ce{O2} can form (up to $\sim$10$^{-6}$) 
depending on the trajectory paths (left panels), and 
(ii) \ce{O2} ice trapped within \ce{H2O} ice does not efficiently form en route into 
the disk (middle panels).
Given that most elemental oxygen is in ices (\ce{H2O} and \ce{CO}) 
at the onset of collapse, gaseous \ce{O2} forms
through photodissociation/desorption of \ce{H2O} ice by stellar UV photons 
in the warm ($>$20 K) protostellar envelope, followed by 
subsequent gas-phase reactions. 
The majority of parcels in each disk have a low final \ce{O2}/\ce{H2O}
ice ratio, $\ll 10^{-2}$.   
However, the upper layers of the disk do have several parcels with a
\ce{O2}/\ce{H2O} ice  ratio higher than $10^{-2}$.
Analysis of the ice composition shows that the 
\ce{O2} ice is associated with \ce{CO2} ice rather than with \ce{H2O}.   
Upon water ice photodissociation, the warm temperatures encountered through 
the protostellar envelope mean that  
\ce{CO2} ice (re)formation is more favorable than that for \ce{H2O} ice.  

We also considered a case where the simulations begin with a similar
fraction of \ce{O2} ice to what is observed in comet 67P/C-G (5\%
relative to water ice). The \ce{O2}/\ce{H2O} ratio throughout both
disks is largely preserved.   
Hence, \ce{O2} which has a prestellar or molecular cloud origin, 
is able to survive the chemical processing en route into the comet-forming 
regions of protoplanetary disks.  However, only little \ce{O2} is
formed during disk formation.

\subsection{Luminosity outburst origin?}
Observational and theoretical studies suggest that the luminosity
evolution of low-mass stars is highly variable, with
frequent and strong eruptive bursts, followed by long periods of
relative quiescence (e.g. \cite[Hartmann \& Kenyon 1985,
Vorobyov et al. 2015]{hartmann1985,vorobyov2015}).
Such luminosity outbursts could have a strong impact on the morphology
and the chemical composition of ices near the protoplanetary disk
midplane. 
If the luminosity outburst is sufficiently strong, warm gas-phase
formation of molecular oxygen could be triggered  by the evaporation
of water ice, if the peak temperature during the outburst  
is higher than $\sim 100$~K. 
We explored the impact of a series of outbursts events in disks on the
formation and recondensation of \ce{O2},
increasing the temperature from 20 to 100 K, every $10^4$ yr for a
total timescale of $10^5$ yr. The Taquet astrochemical 
model described previously was used. 
Initial ice abundances are the median values derived by \cite{oberg2011} 
from interstellar ice observations towards low-mass protostars. 
Thus it is assumed that the ice mantles are initially poor in \ce{O2}.  
The pre-outburst and post-outburst temperature is set to 20~K.
Protoplanetary disk models suggest that the corresponding 
midplane density at this point is $\sim 10^8$ cm$^{-3}$ 
(e.g. \cite[Walsh et al. 2014]{walsh2014}).

We varied several parameters that are thought to impact the gas phase formation of
\ce{O2} during the outburst and the efficiency of recondensiation during the
cooling, such as the grain size, the cosmic ray ionisation rate,
the cooling timescale after the outburst or the peak temperature.
It is found that the maximum amount of \ce{O2} formed during
luminosity outbursts and then trapped within the ice mantle during the
cooling does depend on the explored parameters but never exceeds $\sim
0.1$\% w.r.t. H$_2$O ice. This suggest that luminosity outbursts are too short to
significantly produce \ce{O2} with quantities similar to those
observed in comet 67P/C-G.
Assuming an initial \ce{O2} abundance of $5$\% relative to water ice
results in efficient trapping of \ce{O2} within the water-ice
mantle due to the fast cooling after the outburst. 
However, in that case also other volatile species, such as CO and \ce{N2}, 
become trapped, which is in contradiction with observations towards
67P/C-G.


\section{Conclusions}

The models presented here favour the scenario that molecular 
oxygen in 67P/C-G has a primordial origin (i.e., formed in the parent molecular cloud) 
and has survived transport through the protostellar envelope and into the 
comet-forming regions of protoplanetary disks.
{The ``primordial'' origin of \ce{O2} is in good agreement with
  the conclusions of \cite{mousis2016}. 
  \cite{mousis2016} invoked radiolysis to efficiently convert water
  ice to \ce{O2}. However, we find here that the entrapment and strong association
with water ice combined with low abundance of species like \ce{H2O2},
\ce{HO2}, or \ce{O3} can be explained by an
efficient \ce{O2} formation at the surface of interstellar ices
through oxygen atom recombination in relatively warmer ($\sim 20$ K)
and denser ($n_{\textrm{H}} \gtrsim 10^5$ cm$^{-3}$) conditions than
usually expected in dark clouds.} 
The { weak correlation} of CO and \ce{N2} with
water seen in 67P/C-G is explained by a later formation and freeze-out
of these species in dark clouds with respect to \ce{O2} and water. 
This picture would therefore be consistent with the physical and chemical
properties of our Solar System, such as the presence of short-lived radio isotopes in
meteorites or the orbits of Solar System planets, which suggests that our
Solar System was born in a dense cluster of stars (see \cite[Adams 2010]{adams2010}).

\begin{discussion}

\end{discussion}

\end{document}